\documentclass[aps,prc, nofootinbib,
preprintnumbers,showkeys, reprint,
superscriptaddress]{revtex4-1}

\usepackage{graphicx}
\usepackage{tikz}
\usetikzlibrary{shapes.misc}

\tikzset{
    cross/.pic = {
    \draw[rotate = 45, very thick] (-#1,0) -- (#1,0);
    \draw[rotate = 45, very thick] (0,-#1) -- (0, #1);
    }
}
\usepackage{physics}
\usetikzlibrary{decorations.markings}
\usepackage{subcaption}
\usepackage{dcolumn}
\usepackage{bm}
\usepackage{amsmath}
\usepackage{wasysym}
\usepackage{float}
\usepackage{multirow}
\usepackage{xcolor}
\usepackage{mathtools}
\usepackage{scrextend}
\usepackage[colorlinks=true, 
pdfstartview=FitV, 
bookmarks=true, 
bookmarksnumbered=true, 
breaklinks]{hyperref}
\usepackage{color}
\usepackage[normalem]{ulem} 
\definecolor{blue}{rgb}{0.0, 0.0, 1.0}
\definecolor{red}{rgb}{1.0, 0.0, 0.0}
\definecolor{royalblue}{rgb}{0.0, 0.14, 0.4}
\hypersetup{linkcolor=royalblue, 
	citecolor=blue, 
	urlcolor=royalblue}

\def\orcid#1{\kern .08em\href{https://orcid.org/#1}{\includegraphics[keepaspectratio,width=0.7em]{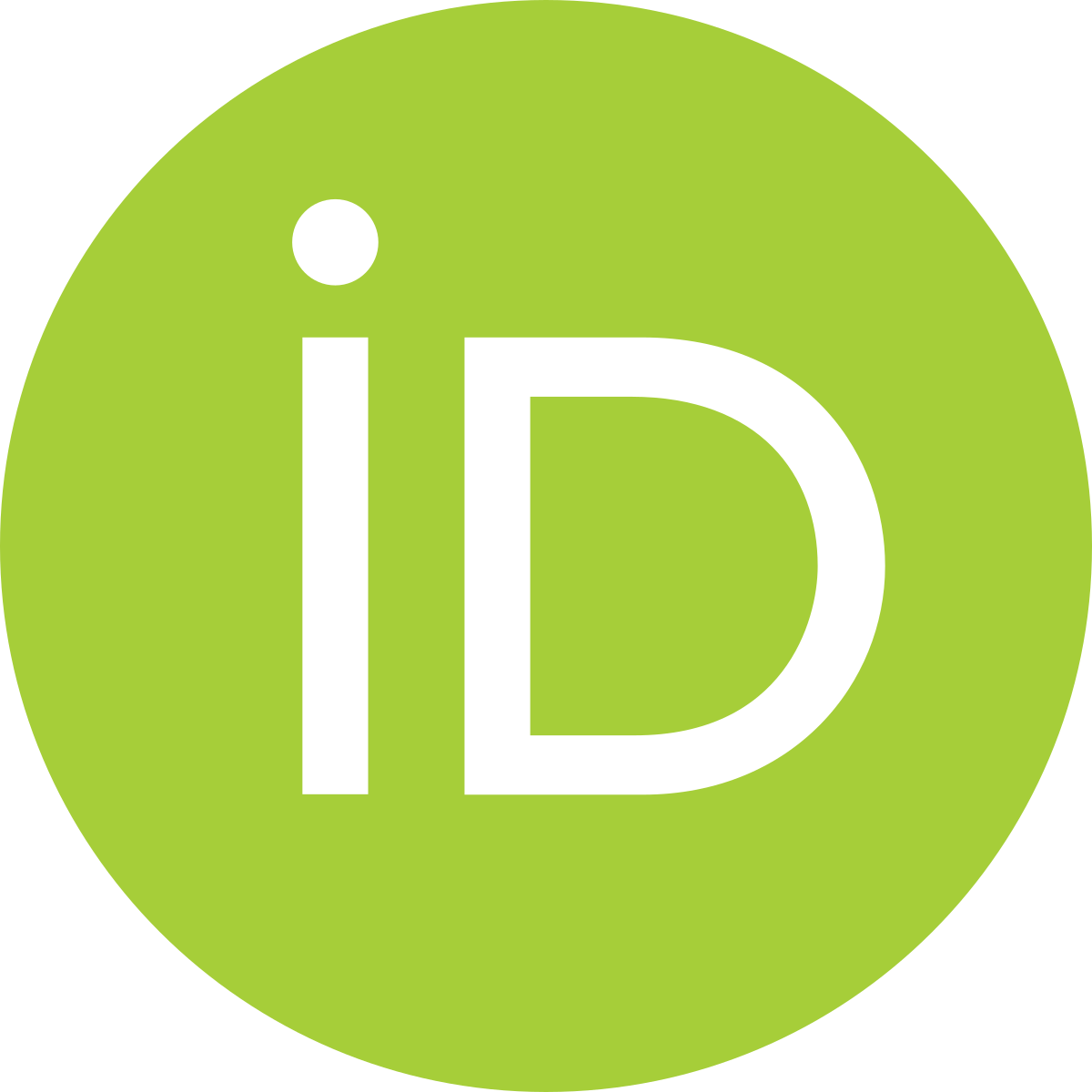}}}

\definecolor{ttcolor}{RGB}{240,248,255}
\definecolor{btcolor}{RGB}{144,238,144}
\definecolor{tbcolor}{RGB}{216,191,216}
\definecolor{bbcolor}{RGB}{255,228,225}

\tikzset{
  >=latex, 
  node/.style={thick,circle,draw=myblue,minimum size=22,inner sep=0.5,outer sep=0.6},
  node in/.style={node,green!20!black,draw=mygreen!30!black,fill=mygreen!25},
  node hidden/.style={node,blue!20!black,draw=myblue!30!black,fill=myblue!20},
  node convol/.style={node,orange!20!black,draw=myorange!30!black,fill=myorange!20},
  node out/.style={node,red!20!black,draw=myred!30!black,fill=myred!20},
  connect/.style={thick,mydarkblue}, 
  connect arrow/.style={-{Latex[length=4,width=3.5]},thick,mydarkblue,shorten <=0.5,shorten >=1},
  node 1/.style={node in}, 
  node 2/.style={node hidden},
  node 3/.style={node out}
}

\begin{document}
\title{Pole structure of $P_{\psi}^{N}(4312)^+$ via machine learning and uniformized $S$-matrix}

\author{Leonarc Michelle Santos\orcid{0000-0003-0444-5544}}
\email[]{lsantos@up.edu.ph}
\affiliation{National Institute of Physics, University of the Philippines Diliman, Quezon City 1101, Philippines}

\author{Vince Angelo A. Chavez\orcid{0009-0009-2373-1985}}
\email[]{vachavez@up.edu.ph}
\affiliation{National Institute of Physics, University of the Philippines Diliman, Quezon City 1101, Philippines}

\author{Denny Lane B. Sombillo\orcid{0000-0001-9357-7236}}
\email[]{dbsombillo@up.edu.ph}
\affiliation{National Institute of Physics, University of the Philippines Diliman, Quezon City 1101, Philippines}

\date{\today}
\begin{abstract}
    \noindent
    We probed the pole structure of the $P_{\psi}^{N}(4312)^+$ using a trained deep neural network. The training dataset was generated using uniformized independent S-matrix poles to ensure that the obtained interpretation is as model-independent as possible. To prevent possible ambiguity in the interpretation of the pole structure, we included the contribution from the off-diagonal element of the S-matrix. Five out of the six neural networks we trained favor $P_{\psi}^{N}(4312)^+$ as possibly having a three-pole structure, with one pole on each of the unphysical sheets—a first in its report. The two poles can be associated to a pole-shadow pair which is a characteristic of a true resonance. On the other hand, the last pole is most likely associated with the coupled-channel effect. The combined effect of these poles produced a peak below the $\Sigma^{+}_C\bar{D}^0$ which mimic the line shape of a hadronic molecule.
    
\end{abstract}		

\maketitle
\section{Introduction}
The idea of a pentaquark was proposed as early as 1964 by Gell-Mann \cite{Gell-Mann1964} and Zweig \cite{zweig1964_3}. However, it was only in 2015, during LHCb Run $1$ \cite{LHCB2015}, that convincing experimental evidence of a pentaquark emerged. In particular, resonances in the $J/\psi p$ invariant mass spectrum were observed, composed of $c\bar{c}uud$ quarks \cite{LHCB2015}, leading to the discovery of two such $P_c$ states in 2015: the $P_c(4380)^+$ and $P_c(4450)^+$.

With the improved statistics of LHCb Run $2$ \cite{Pc2019,LHCB2019hepdata.89271}, the structure seen in the $P_c(4450)^+$ from Run $1$ was resolved into two narrow states: $P_c(4440)^+$ and $P_c(4457)^+$. Additionally, a new narrow state with $7.3\sigma$ significance was discovered: the $P_c(4312)^+$ pentaquark, now referred to as $P_\psi^N(4312)^+$. The $P_\psi^N(4312)^+$ is of particular interest due to its relatively clean signal.

Due to the proximity of $P_\psi^N(4312)^+$ to the $\Sigma_c\Bar{D}^0$ threshold, which lies roughly $5$ MeV below it, several studies have adopted a molecular interpretation for $P_\psi^N(4312)^+$ \cite{Guo_oller2019, oset2019, QCDsumrules, zhu_one_boson2019,He2019,molecular_scenario_chen_2019,KinematicsPhysRevLett.124.072001,virtualLHCbcollab}. A bottom-up approach in Ref.~\cite{Pc4312} supports this molecular interpretation, complementing the top-down analyses. However, additional top-down studies, such as those in Ref.~\cite{ALI2019365compact} and Ref.~\cite{Nakamura2021correct}, argue in favor of a compact structure and a kinematic origin, respectively, to reproduce the experimental observations.

Meanwhile, applications of machine learning in hadron spectroscopy were first developed in Ref.~\cite{sombilloPhysRevD.102.016024, Sombillo2021classfyng}, initially used for a single-channel analysis of nucleon-nucleon scattering. This approach was later generalized for a two-channel analysis in Ref.~\cite{SombilloPhysRevD.104.036001}. Soon after, machine learning was applied to probe the $P_\psi^N(4312)^+$ pentaquark in Refs.~\cite{DeepLearningExHad, ZHANG2023981, co2024deep}.

Ref.~\cite{DeepLearningExHad} extended the work in Ref.~\cite{Pc4312} by using a deep neural network to classify whether $P_\psi^N(4312)^+$ is a bound state or a virtual state and on what sheet its pole is located. This work favored a virtual interpretation, locating it on the $[tb]$ Riemann sheet, supporting the findings in Ref.~\cite{Pc4312}.

Ref.~\cite{ZHANG2023981}, on the other hand, employed a neural network to determine the quantum numbers of the $P_c$ states within a pionless effective field theory framework. Their neural network successfully distinguished the quantum numbers of $P_c(4440)$ and $P_c(4457)$, which the normal $\chi^2$ fitting approach failed to do. However, the authors acknowledged that these results should be interpreted cautiously. A one-pion exchange potential is necessary to make a definitive statement about hadronic molecules.

Lastly, Ref.~\cite{co2024deep} used machine learning to investigate the plausible triangle singularity picture raised in Ref.~\cite{Nakamura2021correct}. Their findings ruled out the triangle diagrams they considered and suggested that the $P_\psi^N(4312)^+$ pentaquark is likely a molecular structure, favoring a single pole on the $[bt]$ sheet.

As we can observe, machine learning and neural networks can be tailored to meet our specific objectives. In particular, in Refs.~\cite{DeepLearningExHad, ZHANG2023981, co2024deep}, machine learning was applied for a classification task across different frameworks. The aim of this paper is to build upon what was emphasized in Ref.~\cite{co2024deep}, using neural networks as a model selection tool. In contrast to previous studies on $P_\psi^N(4312)^+$ utilizing neural networks, we used a uniformized $S$-matrix to construct our training dataset, which comprised eight classes corresponding to eight different pole configurations. Moreover, we incorporated the inelastic contribution of the $\Lambda_b^0 \to J/\psi p$ decay into our full $S$-matrix. Our earlier work in Ref.~\cite{SantosSombilloPhysRevC.108.045204} showed that excluding the $S_{12}$ contribution from the full $S$-matrix can lead to ambiguous line shapes, potentially obscuring hidden physics. With this, six neural network models were trained using our generated training dataset and evaluated using our validation dataset for accuracy, macro F1 scores, and the recall and precision scores of each class. Our neural networks inferred that the pole structure of $P_\psi^N(4312)^+$ may be a $3$-pole structure consisting of a pole on each of the three unphysical sheets of the $\Lambda_b^0 \to J\psi/p$ decay -- a first report of this pole structure.



The content of this paper is organized as follows: In Section~\ref{sec:II}, we show how we constructed an $S$-matrix with independent poles using uniformization introduced in Refs.~\cite{Newton, Kato1965, Yamada2020, Yamada2021, yamada2022near}. In Section~\ref{sec:III}, we discuss how we generated our training dataset, the architecture of our neural networks, and their performance against a validating dataset. We discuss the inference results in Section~\ref{sec:IV} and its possible physical interpretation. We conclude and present an outlook for future works in Section~\ref{sec:V}.

\section{Independent $S$-matrix poles via uniformization}\label{sec:II}

The elements of a two-channel $S$-matrix are given by
\begin{equation}\label{eq:Sdiag}
S_{11}(p_1,p_2)=
\dfrac{D(-p_1,p_2)}
{D(p_1, p_2)}; \quad S_{22}(p_1,p_2)=
\dfrac{D(p_1,-p_2)}
{D(p_1, p_2)},
\end{equation}
and
\begin{equation}\label{eq:Soffdiag}
S_{12}^2=S_{11}S_{22}-
\text{det}S.
\end{equation}
where $D(p_1,p_2)$ is the Jost function \cite{LeCouter1960, Newton1961, Newton1962RelaxCons,Kharakhan1971}. The subscripts correspond to the channel index with $1$ representing the lower mass channel, and $2$ for the higher mass channel. When mapped to the complex energy plane, the full $2\times2$ $S$-matrix possesses two branch points, corresponding to the two channel threshold energies $\epsilon_1$ and $\epsilon_2$, and in general, may possess pole singularities. 

The branch cuts are chosen to run along the positive real energy axis, with two branch points opening up four Riemann sheets \cite{Newton, Rakityansky2022}. In this work, we adopted the notation of Ref.~\cite{PearceGibson} in labeling our Riemann sheets. The notation is $[XY]$, where $X$ corresponds to the sheet of the first channel and $Y$ to the sheet of the second channel. The strings $X$ and $Y$ can be $t$ or $b$ denoting the top sheet or bottom sheet, respectively. The correspondence of Pearce and Gibson’s notations with the much more used Frazer and Hendry’s \cite{Frazer1964}, is $[tt] \to I$, $[bt] \to II$, $[bb] \to III$, and $[tb] \to IV$.

The other singularity of the $S$-matrix are its poles, manifesting from the zeros of the Jost function $D(p_1,p_2)$. By extending and connecting the analytic region of the Jost functions in the denominator and numerator of equations \eqref{eq:Sdiag} and \eqref{eq:Soffdiag}, we establish the correspondence between the simple poles of the $S$-matrix and quantum states \cite{Newton, Taylor}. Its state correspondence can be a bound state, virtual state, or a resonance, depending on what Riemann sheet the poles are located on and their numbers \cite{Badalyan1982, MorganPennington1991, Morgan1992, MorganPennington1993}.

Having these said, we can perform a bottom-up analysis by parametrizing the $S$-matrix to fit into the scattering cross section $\mathrm{d}\sigma/\mathrm{d}\Omega$. From the parametrization, we look at the analytic properties of the resulting $S$-matrix and infer the physics from it. Moreover, we only need three constraints on the $S$-matrix: (1) analyticity, (2) hermiticity, and (3) unitarity below the threshold \cite{Newton, Taylor}.

In this work, we parametrized our $S$-matrix by using the uniformized variable $\omega$. The uniformization scheme was introduced in Refs.~\cite{Newton, Kato1965} and further explored in Refs.~\cite{Yamada2021,Yamada2020}. We proceed as follows.

From the expression of the $k$th channel's momentum in the two-hadron center-of-mass frame given by
\begin{equation}
	p_k^2=\dfrac{(s-\epsilon_k^2)
		\left[s-\epsilon_k(\epsilon_k-4\mu_k)\right]}
	{4s}
	\label{eq:actual_p}
\end{equation}
where $\epsilon_k$ and $\mu_k$ are the threshold energy and reduced mass of the $k$th channel, we write the invariant Mandelstam variable $s$ as
\begin{equation}
	s = \epsilon_k^2 + \dfrac{\epsilon_k}{\mu_k}|\vec{p}_k|^2
	\left[1+\mathcal{O}\left(\dfrac{|\vec{p}_k|^2}{\epsilon_k^2}\right)\right]=\epsilon_k^2+q_k^2.
	\label{eq:invariant_s}
\end{equation}
Here we introduced the new momentum variable $q_k$ to simplify our scaling. We define the uniformized variable $\omega$ by the transformation
\begin{equation}
	\omega=\dfrac{q_1+q_2}
	{\sqrt{\epsilon_2^2-\epsilon_1^2}};
		\quad\quad
		\dfrac{1}{\omega}=
		\dfrac{q_1-q_2}{\sqrt{\epsilon_2^2-\epsilon_1^2}}.
			\label{eq:uniformization}	
\end{equation}

\begin{figure}[tbhp!]
    \centering
    \includegraphics[scale = 0.35]{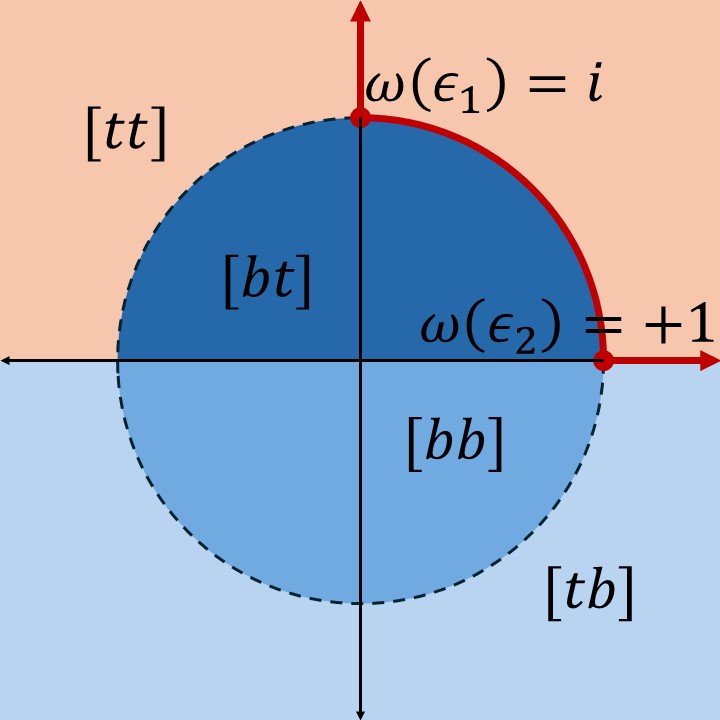}
    \caption{The mapping of the four Riemann sheets of the complex energy $E$ into the $\omega$-plane. The red line was once the branch cuts of $\epsilon_1$ and $\epsilon_2$.}
    \label{fig:omegaplane}
\end{figure}

With this transformation, our two-variable $S$-matrix is reduced into a single variable $S$-matrix. Furthermore, the branch point singularity of $S(p_1,p_2)$ is removed due to the linear dependence of the $\omega$ with $(q_1,q_2)$. We can safely characterize near-threshold peaks through uniformization, as it properly and rigorously incorporates resonances and threshold behaviours \cite{Yamada2020}. With $\omega$, the four Riemann sheets of the complex energy are reduced to a single complex $\omega$-plane. We show the mapping of the four $[XY]$ Riemann sheets to the $\omega$-plane in Figure~\ref{fig:omegaplane}. The detailed description of such mapping can be accessed in Refs.~\cite{Kato1965,yamada2022near}.

With the defined uniformized variable $\omega$ in Eq.~\ref{eq:uniformization}, we could express the two-variable Jost function $D(p_1,p_2)$ as a single variable of $\omega$, $D(\omega)$. Given this, the matrix elements of the two-channel $S$-matrix takes the form
\begin{align}\label{eq:smatrix}
	\begin{split}
	    S_{11}(\omega)=\dfrac{D(-1/\omega)}{D(\omega)};& \quad \quad
			S_{22}(\omega)=\dfrac{D(1/\omega)}{D(\omega)}; \\
			\text{det}(S)&=\dfrac{D(-\omega)}{D(\omega)}.
	\end{split} 
\end{align}
The line shape amplitude is constructed from the $T$-matrix. This is extracted from the $S$-matrix via the relation $S_{j,k}=\delta_{j,k}-2iT_{j,k}$ where $\delta_{j,k}$ is the Kronecker delta.

To construct an $S$-matrix with a pole at $\omega = \omega_{\text{pole}}$, we can use a polynomial Jost function expressed as
\begin{equation}
	D(\omega)=\dfrac{(\omega-\omega_\text{pole})
	(\omega+\omega_\text{pole}^*)
	(\omega-\omega_\text{reg})
	(\omega+\omega_\text{reg}^*)}{\left(\omega_\text{pole}\omega_\text{reg}\right)^2}
	.
	\label{eq:jost}
\end{equation}
Several factors were introduced to satisfy the constraints of the $S$-matrix: The factor $\left(\omega_\text{pole}\omega_\text{reg}\right)^{-2}$ warrants unitarity below the threshold. The negative conjugate terms were introduced to satisfy the hermiticity of the $S$-matrix below the lowest threshold \cite{Yamada2020,Taylor}. The factor $\omega_{\text{reg.}}$, called the pole regulator, is an extra term added to ensure the analyticity of the $S$-matrix. Specifically, the regulator ensures that $S_{kk}(\omega) \to 1 $ as $\omega \to \infty$ \cite{Kato1965,Newton}. To ensure that only the $\omega_{\text{pole}}$ term is relevant to the line shape amplitude, we parametrize the pole regulator as $\omega_{\text{reg}} = e^{-i\pi/2}/|\omega_{\text{pole}}|$. Looking at Figure~\ref{fig:omegaplane}, we see that the phase factor casts the regulator either on the $[bb]$ or $[tb]$ sheet, far away from the scattering region. 

Generalizing Jost function \eqref{eq:jost} to construct an $S$-matrix with $N$-poles, we can simply compose the Jost function and write it as
\begin{eqnarray}\label{eqn:Jostcompose}
	D(\omega)&=
	\prod_{\{t=\text{pole}_n,\text{reg}_n\}}^{N}\omega_t^{-2}
	(\omega-\omega_t)
    (\omega+\omega_t^*).
	\label{eq:bigjost}
\end{eqnarray}
With Eq.~\eqref{eqn:Jostcompose}, the resulting $S$-matrix will have poles independent from each other. Moreover, we can easily control the poles through the single variable $\omega_{\text{pole}}$ without worrying about the constraints of the $S$-matrix, since their Riemann sheet locations are implicitly embedded in $q_1^{(N)}$ and $q_2^{(N)}$.

This pole independent $S$-matrix framework is more suitable for our work as compared to other parametrization schemes such as the Flatté or effective range expansion. In these parametrizations, upon fixing one of the poles, the position of the other poles will necessarily be constrained. As an example, it was demonstrated in Ref.~\cite{Frazer1964} that a specific coupled-channel effective range expansion can never produce poles on the $[bb]$ sheet. Moreover, although the Flatté parametrization or the effective range expansion can be constructed without any reference to a model, one can always find an effective coupled-channel potential that can reproduce its pole trajectories \cite{PearceGibson, Frazer1964,Hanhart2014,Hyodo:2014bda}. This implies that these parametrizations implicitly favors already a given pole trajectory. With our independent poles parametrization, we can cover a wider model space of poles without violating the constraints of the $S$-matrix. This also helps us further our idea of designing a neural network with as few biases as possible.

We note that the parametrization scheme discussed in this section is not new. The uniformization transformation \eqref{eq:uniformization} and Jost function \eqref{eq:jost} were first introduced in 1965 by Kato \cite{Kato1965}. We conclude this section by recalling an important result from our previous work in Ref.~\cite{SantosSombilloPhysRevC.108.045204}. With the parametrization scheme discussed herein, cases of ambiguous line shape amplitudes arise in the $S_{11}$ channel. To resolve this ambiguity, one must examine the $S_{12}$ channel and incorporate its contribution into the full $S$-matrix, i.e., we must use $S \approx S_{11} + S_{12}$ instead of $S \approx S_{11}$. Having addressed this, we will now proceed to the next section, where we discuss the framework of our neural network.

\section{Machine learning formalism}\label{sec:III}
The main goal of our work is to apply machine learning to determine the plausible pole structure of the $P_\psi^N(4312)^+$ signal in the $J/\psi p$ invariant mass spectrum. To achieve this, the work is divided into four stages: (1) constructing the training dataset, (2) designing neural network models, (3) training and validation, and finally, (4) inferring the experimental data itself. It can be argued that the most crucial part is the construction of the training dataset, as the quality of the neural network depends on it. However, designing and tuning the neural network architecture, parameters, and hyperparameters are also essential for achieving satisfactory performance. In this work, we emphasize the construction of a quality training dataset, while considering finetuning the neural network models in future work. We first discuss how we generated our training dataset, followed by an explanation of our neural network design. We then elaborate on curriculum learning, a method employed to help our model learn efficiently. Finally, we validate the models using a validation dataset. The inference stage is discussed in the next section.




\subsection{Generation of training dataset}
Our training dataset is a tuple consisting of the input and output datasets. The output data consists of eight plausible pole structures of the $P_\psi^N(4312)^+$, and the input data comprises the energy range and the corresponding line shape amplitude of the pole structures. For this study, we considered eight possible pole configurations, as listed in Table~\ref{tab:pole_label}.
\begin{table}[tbhp!]
\caption{\label{tab:pole_label}%
Classification output-node label.
}
\begin{ruledtabular}
\begin{tabular}{l l dr}
\textrm{Class label} & \textrm{$S$-matrix pole configuration} \\
\colrule
$0$ & $1$ pole on $[bt]$ \\ 
$1$ & $1$ pole on $[bb]$ \\ 
$2$ & $1$ poles on $[tb]$ \\
\\
$3$ & $1$ pole on $[bt]$ and $1$ pole on $[bb]$ \\
$4$ & $1$ pole on $[bb]$ and $1$ pole on $[tb]$ \\
\\
$5$ & $1$ pole on $[bb]$, $1$ pole on $[tb]$, $1$ pole on $[bt]$ \\
$6$ & $2$ pole on $[bb]$, $1$ pole on $[tb]$ \\
$7$ & $1$ pole on $[bb]$, $2$ pole on $[tb]$ \\
\end{tabular}
\end{ruledtabular}
\end{table}

Labels $0$ through $2$ correspond to a one-pole configuration. These labels may represent an unstable bound state, an inelastic virtual state, a Breit-Wigner pole, or a coupled-channel pole, depending on their half-plane location and Riemann sheet \cite{Badalyan1982}. Generally, they correspond to a molecular state, supporting interpretations found in Refs.\cite{Guo_oller2019, oset2019, QCDsumrules, zhu_one_boson2019,He2019,molecular_scenario_chen_2019,KinematicsPhysRevLett.124.072001,virtualLHCbcollab}. On the other hand, labels $3$ and $4$ denote a two-pole configuration indicative of a compact state \cite{MorganPennington1991,Morgan1992}, supporting the findings of Ref.\cite{ALI2019365compact}. Finally, labels $5$ through $7$ were included to account for the ambiguous line shapes on the elastic channel, as discussed in Ref.\cite{SantosSombilloPhysRevC.108.045204}. While a larger model space would be ideal, considering this is our first attempt using uniformization and incorporating the inelastic contribution into the full $S$-matrix, the eight pole configurations listed in Table\ref{tab:pole_label} will suffice.

We considered the limited energy region $[4212,4412]$ MeV for practicality and to isolate the other resonance peaks. Referring to the 2019 LHCb Run 2 data, $100$ data points appear within this vicinity \cite{virtualLHCbcollab,LHCB2019hepdata.89271}. Given this, the energy axis of our line shape amplitudes must comprise $100$ data points as well. Hence, we divided the energy region $[4212,4412]$ MeV into 100 equally spaced bins. A random energy point is chosen from each bin using a uniform probability distribution. The collection of these random energy points from each bin comprises the energy axis. With this scheme, we can interpret each bin as the energy resolution of the particle detector. 

The lineshape amplitude is computed from the random energy points using \cite{Frazer1964, Pc4312}
\begin{equation}\label{eqn:finalamplitude}
f\left(\sqrt{s}\right) = \rho\left(\sqrt{s}\right)\left[\abs{T\left(\sqrt{s}\right)}^2+b\left(\sqrt{s}\right)\right].
\end{equation}
Here, $\rho$ represents the phase space, and $b\left(\sqrt{s}\right)$ is a quadratic polynomial used to simulate background noise and capture the tail behavior of the amplitude in the energy region $[4212,4412]$ MeV. The $T$-matrix is defined as
\begin{equation}
T\left(\sqrt{s}\right) = T_{11}\left(\sqrt{s}\right) + T_{12}\left(\sqrt{s}\right),
\end{equation}
where we include the off-diagonal term to enable the neural network to distinguish ambiguous line shapes \cite{SantosSombilloPhysRevC.108.045204}. It is important to note that the parametrization \eqref{eqn:finalamplitude} assumes dominance of the $s$-wave due to the strong cusp at the $\Sigma_c^+\Bar{D}$ threshold.

The poles, generated by the zeros of the Jost function \eqref{eqn:Jostcompose}, are constrained within the vicinity of the $\Sigma_c^+\Bar{D}$ threshold. Specifically, the randomly generated poles have real and imaginary parts within the range
\begin{align*}
    \begin{dcases}
    \epsilon_2 - 100  &\leq \text{Re} \ E_{\text{pole}} \leq \epsilon_2 + 100   \\
    0.5 &\leq \abs{\text{Im}\ E_{\text{pole}}} \leq 50
\end{dcases}
\end{align*}

Overall, our training data $x^j$ consists of $100$ data points for the energy axis and the corresponding $100$ data points for the line shape amplitudes \eqref{eqn:finalamplitude}, resulting in a total of $200$ points for $x^j$. Including the energy axis provides an additional feature for the neural network to distinguish.

In summary, the training dataset $(x^j,y^j)$ is a tuple consisting of the input $x^j$, which comprises the energy and corresponding amplitude, and the output $y^j$, corresponding to the eight pole configurations listed in Table~\ref{tab:pole_label}. We generated $2500$ instances of each pole configuration for each curriculum (defined in Sec.~\ref{subsec:currmethod}). For every label generated, $0.80$ of it is allocated for training and $0.20$ for testing.

\subsection{Neural network architecture}
A neural network consists of an input layer, hidden layers, and an output layer. Its main goal is to optimize the parameters $\theta \equiv \left\{W^i_j,b^i\right\}$ of the linear function
\begin{equation}\label{eqn:z_linear}
z_i = W_j^i x^j + b^i,
\end{equation}
where $W_j^i$ is the weight matrix, $x^j$ is the input dataset, and $b^i$ is the bias vector. The weight matrix $W_j^i$ and $b^i$ are initialized, and $z_i$ passes through the hidden layers. Within the hidden layers, the linear function $z_i$ is nonlinearized using the Rectified Linear Unit (ReLU) activation function \cite{relu1, relu2, relu3}
\begin{equation}
\textrm{max}\left(0,z_i\right) = \frac{z_i+|z_i|}{2} = \begin{dcases}
			z_i & \text{if} \ z_i > 0, \\
			0 & \text{otherwise}.
		  \end{dcases}	
\end{equation}
Depending on the input $z_i$, only selected nodes in the hidden layers will be activated. The data then passes through until it reaches the output layer. Nodes in the output layer are equipped with a softmax activation function, defined as \cite{softmaxactivationbook}
\begin{equation}
\text{softmax}(z_n^{(L+1)}) = \frac{\text{exp}(z_n^{(L+1)})}{\sum_m^{N_{L+1}} \text{exp}(z_m^{(L+1)}) },
\end{equation}
where $L$ is the index of the last hidden layer. The data at the output layer corresponds to the probability distribution across the predicted output classes. Only nodes whose probability exceeds our threshold of $1/8 = 0.125$ will be activated, and the label with the highest probability will be selected from the output layer as the label for the input $x^j$.

To obtain the optimal $\theta$ parameters, we utilize a backpropagation algorithm on the cost function $C(\theta)$, given by
\begin{equation}\label{eqn:cost}
C(\theta) = \frac{1}{X}\sum_{x} \Vec{a}(x) \cdot \log \left[\Vec{y}_{\theta}(x)\right].
\end{equation}
Here, the input node values are contained in the array $x$, and $\Vec{a}(x)$ is an array denoting the true label of input $x$. The output node of the neural network is denoted by $\Vec{y}_{\theta}(x)$, and $X$ represents the total number of elements in the training set. The cost function in Eq.~\eqref{eqn:cost} is called the softmax cross-entropy cost function, typically used for a general classification problem \cite{Aggarwal2018}. Generally, the cost function contrasts the predicted label of the input data $x^j$ with its true label $y^j$, and our objective is to minimize $C(\theta)$.

\begin{figure}[tbhp!]
\includegraphics[scale=0.5]{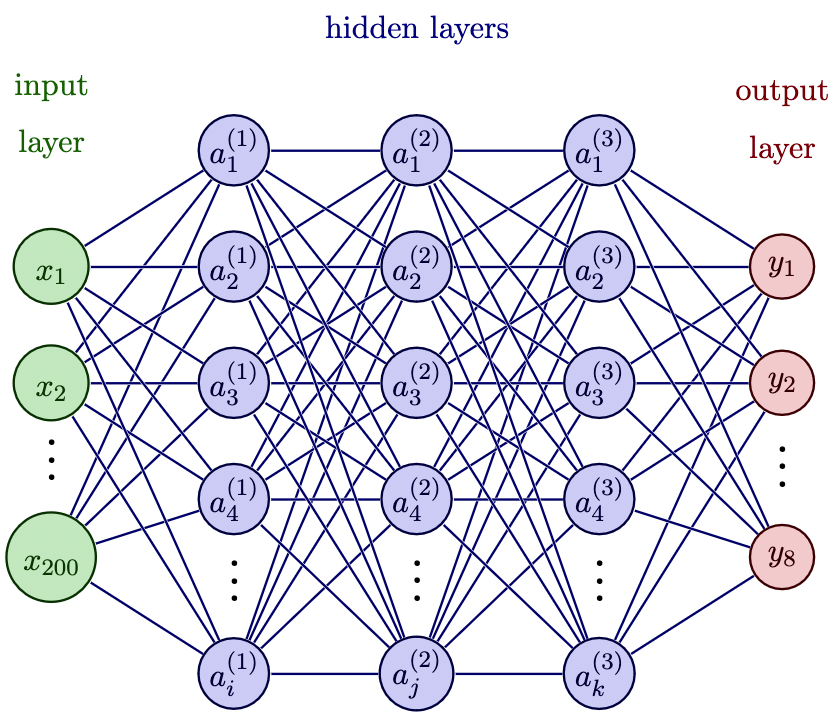}
\caption{\label{fig:DNNschematic} Schematic diagram of the DNN to be trained. TikZ code to produce this figure is from \cite{tikzNN} which is licensed with CC BY-SA 4.0.}
\end{figure}

After minimizing $C(\theta)$ using a backpropagation algorithm, the weight matrix $W^i_j$ and the bias vector $b^i$ are recalibrated accordingly. The process from the initial step to this point is referred to as an epoch. This procedure is repeated for a number of epochs to find the optimal $\Bar{\theta}$. The schematic diagram of our DNN is shown in Figure~\ref{fig:DNNschematic}.

To this end, we considered six neural network architectures, listed in Table~\ref{tab:modellist}. 
\begin{table}[tbhp!]
\centering
\caption{\label{tab:modellist}Model list and their corresponding architecture}
\begin{ruledtabular}
\begin{tabular}{ccdr}
\textrm{Model} & \textrm{Architecture} \\
\colrule
$1$ & $[100-200]$ \\
$2$ & $[200-100]$ \\
$3$ & $[200-200]$ \\
$4$ & $[100-200-300]$ \\
$5$ & $[300-200-100]$ \\
$6$ & $[300-300-300]$ \\
\end{tabular}
\end{ruledtabular}
\end{table}

The notation $[X-Y-Z]$ denotes the number of nodes within each hidden layers. For example, model $1$ with an architecture $[100-200]$ means that it is a neural network with two hidden layers, with $100$ and $200$ nodes, respectively. All architectures are equipped with a \texttt{SMORMS3} optimizer (squared mean over root mean squared cubed) \cite{Funk_2015, optimizer_SMORMS_Chainer}. The batch size was set to $64$ throughout the training and ran for a total of $2000$ epochs. We conducted initial training where the models were fed with all eight labels listed in Table~\ref{tab:pole_label}. This proved futile as all models failed to learn. This issue was also observed in Ref.~\cite{SombilloPhysRevD.104.036001}. To circumvent this problem, we implemented curriculum learning, as discussed in the next subsection.

\subsection{Curriculum method}\label{subsec:currmethod}
Curriculum learning, initially introduced in \cite{ELMAN199371}, emerged as a strategy for training neural networks to grasp intricate relationships and embedded clauses within complex sentences. The concept of "starting small" proved effective in tackling certain challenges, advocating a gradual progression in difficulty over time. As classification problems grew in complexity, a more systematic approach to handling training datasets became imperative. Subsequent research \cite{CurriculumLearning0, CurriculumLearning1, CurriculumLearning2} delved into the refinement and extension of the curriculum learning method. The underlying principle of "starting small" posits that mastering simpler concepts lays a solid groundwork for tackling more complex ones. In the field of hadron spectroscopy, the feasibility of curriculum learning in pole classification was demonstrated in Ref.~\cite{SombilloPhysRevD.104.036001}.

The concept of curriculum learning involves initially training a neural network using a simple dataset and gradually introducing more complex data over time. We organized the dataset into curricula, as listed in Table~\ref{tab:pole_label}.
\begin{table}[tbhp!]
    \centering
    \caption{\label{tab:tablecurr}Curricula arrangement of the label classes.}
    \begin{ruledtabular}
        \begin{tabular}{ccc} 
            \textrm{Curr. label} & \textrm{Included labels} & \textrm{Config. presented} \\
            \colrule
            $1$ & Classes $0,1,2$ & At most $1$ pole\\
            \\
            $2$ & Curr. $1$ $+$ classes $3,4$ & At most $2$ poles \\
            \\
            $3$ & Curr. $3$ $+$ classes $5,6,7$ & At most $3$ poles \\
        \end{tabular}
    \end{ruledtabular}
\end{table}

We organized the classes heuristically; for instance, referring to Table~\ref{tab:pole_label}, labels $0$ to $2$ represent the simplest configurations, followed by labels $3$ and $4$, and finally, the most complex configurations are represented by labels $5$ to $7$. Curriculum $1$ includes only configurations with at most $1$ pole. Curriculum $2$ incorporates both $1$-pole and $2$-pole configurations, while Curriculum $3$ encompasses all configurations of interest.

For each curriculum, we generated a new batch of dataset to mitigate overfitting, with each batch comprising $2,500$ lineshapes per class. For example, Curriculum $1$ encompasses a total of $7,500$ lineshapes, distributed evenly across the three classes, while Curriculum $2$ encompasses a total of $12,500$ lineshapes, evenly distributed across the five classes.

\begin{figure*}[tbhp!]
    \centering
    \includegraphics[scale = 0.3]{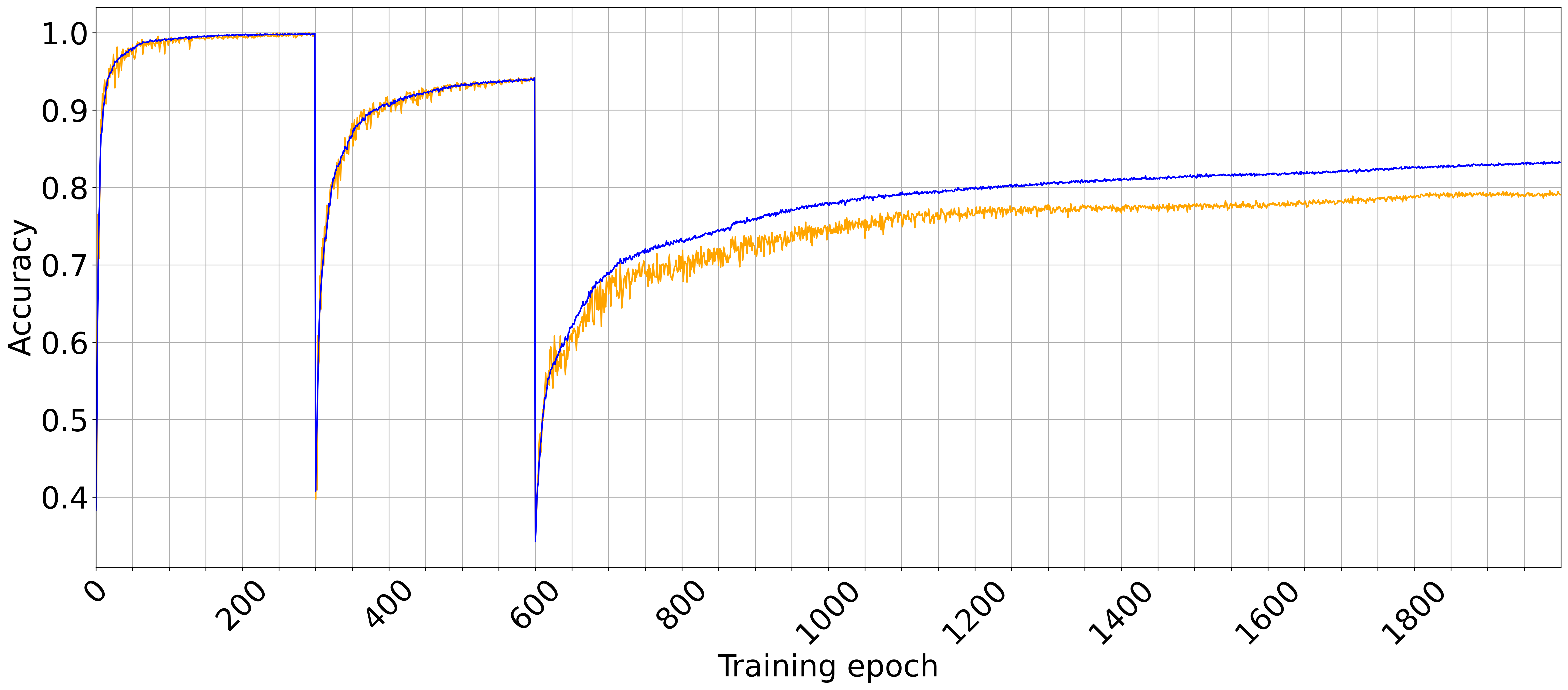}
    \caption{Accuracy vs. epoch of Model $2$: $[200-100]$. The blue line represents the training performance and the orange line the testing performance. The large dips in performance signifies the introduction of the new labels.}
    \label{fig:accu_vs_epoch_model2}
\end{figure*}

After organizing the dataset into curricula, we trained six models using curriculum $1$ for $300$ epochs, followed by curriculum $2$ for another $300$ epochs, and then continued training them with curriculum $3$ for an additional $1400$ epochs, totaling $2000$ epochs across all curricula. The performance of all six models exhibits similar accuracy trends over time. For instance, Figure~\ref{fig:accu_vs_epoch_model2} displays the accuracy evolution of model $2$, showing two significant dips that signify transitions between curricula. These dips occur because new labels are introduced as we progress through the curricula. We summarize the training and testing performance of the six models in Table~\ref{tab:modelperform}.

\begin{table}[tbhp!]
\caption{\label{tab:modelperform}Training and testing accuracy of each model.}
\begin{ruledtabular}
\begin{tabular}{ccc}
\textrm{Model} & \textrm{Training accu.} & \textrm{Testing accu.}\\
\colrule
$1$ & $83\%$ & $79\%$ \\
$2$ & $83\%$ & $79\%$ \\
$3$ & $87\%$ & $81\%$ \\
$4$ & $79\%$ & $75\%$ \\
$5$ & $88\%$ & $82\%$ \\
$6$ & $89\%$ & $83\%$ \\
\end{tabular}
\end{ruledtabular}
\end{table}


The accuracy scores listed in Table~\ref{tab:modelperform} are simply the percentage of the correct predictions of the models over the total training (testing) dataset. Each of the six models achieved a training accuracy of no less than $79\%$ and a testing accuracy of at least $75\%$. However, it is essential to cross-validate these models to assess their performance on unknown datasets. For this purpose, we created a validation dataset comprising all labels, with each label having $2,500$ samples. The subsequent subsection provides a detailed discussion of our models' performance using this validation dataset.

\subsection{Model performance using a validation dataset}
\begin{figure*}[tbhp!]
    \centering 
\begin{subfigure}{0.25\textwidth}
  \includegraphics[width=\linewidth]{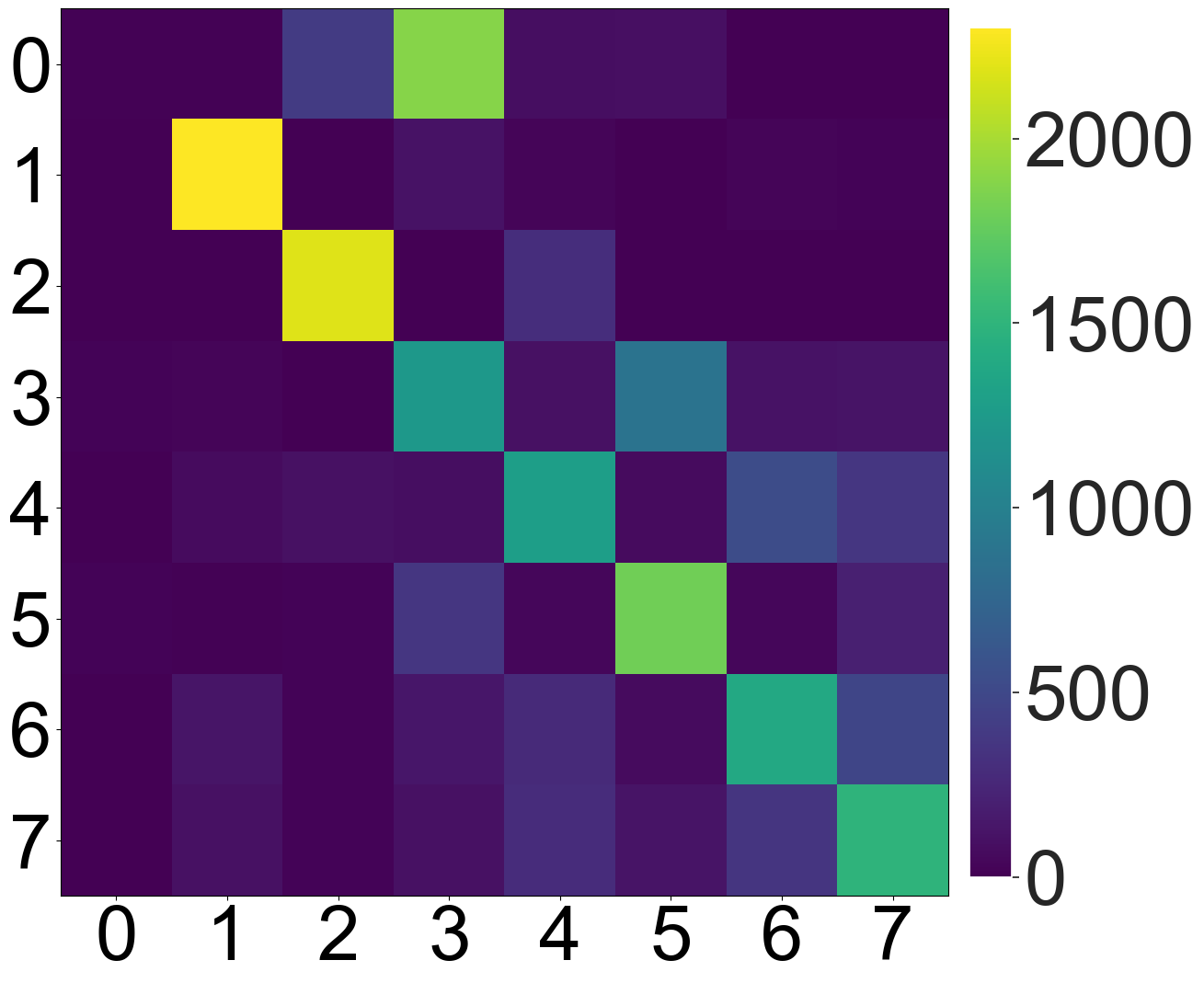}
  \caption{Model 1: $[100-200]$}
  \label{fig:1}
\end{subfigure}\hfil 
\begin{subfigure}{0.25\textwidth}
  \includegraphics[width=\linewidth]{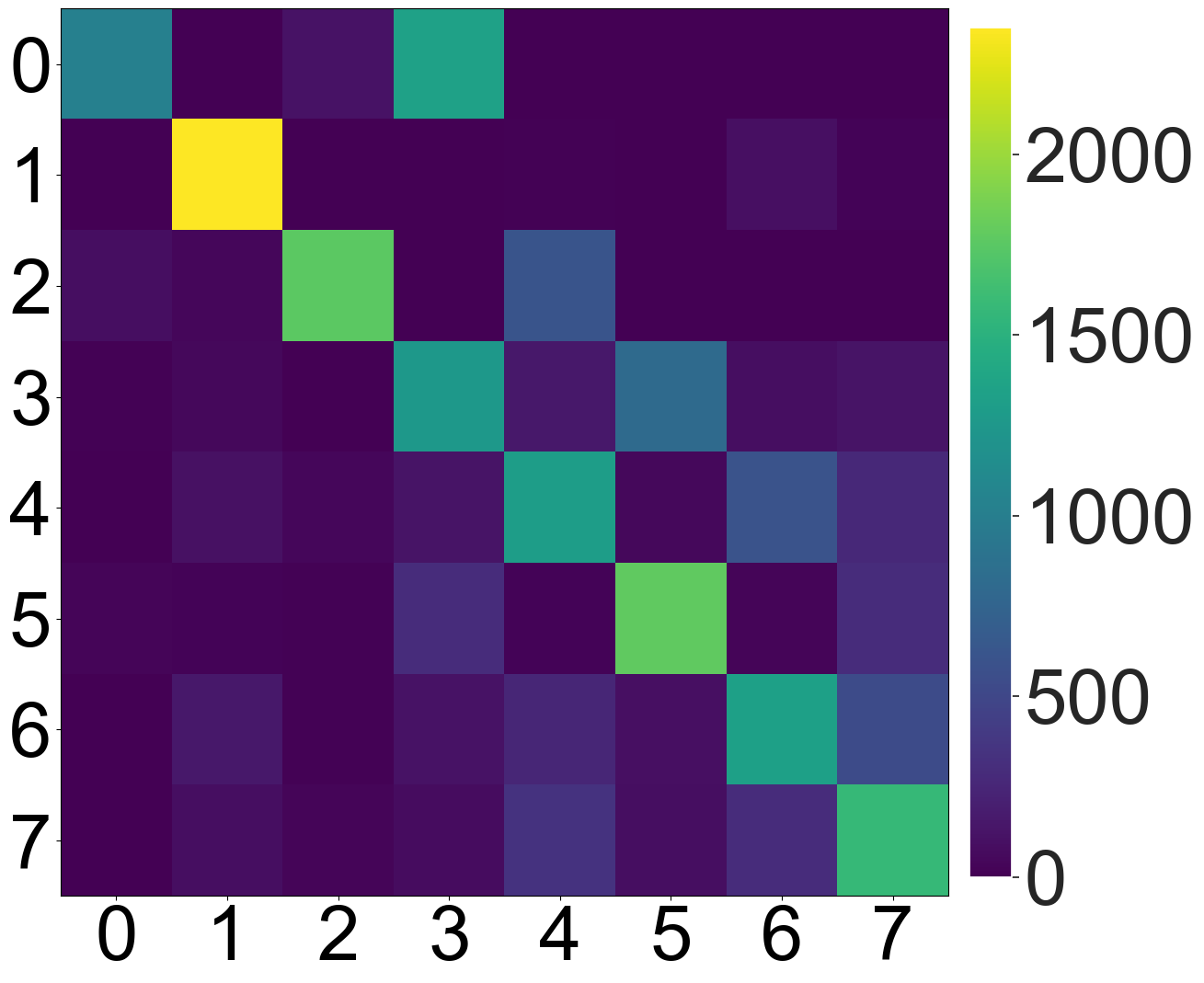}
  \caption{Model 2: $[200-100]$}
  \label{fig:2}
\end{subfigure}\hfil 
\begin{subfigure}{0.25\textwidth}
  \includegraphics[width=\linewidth]{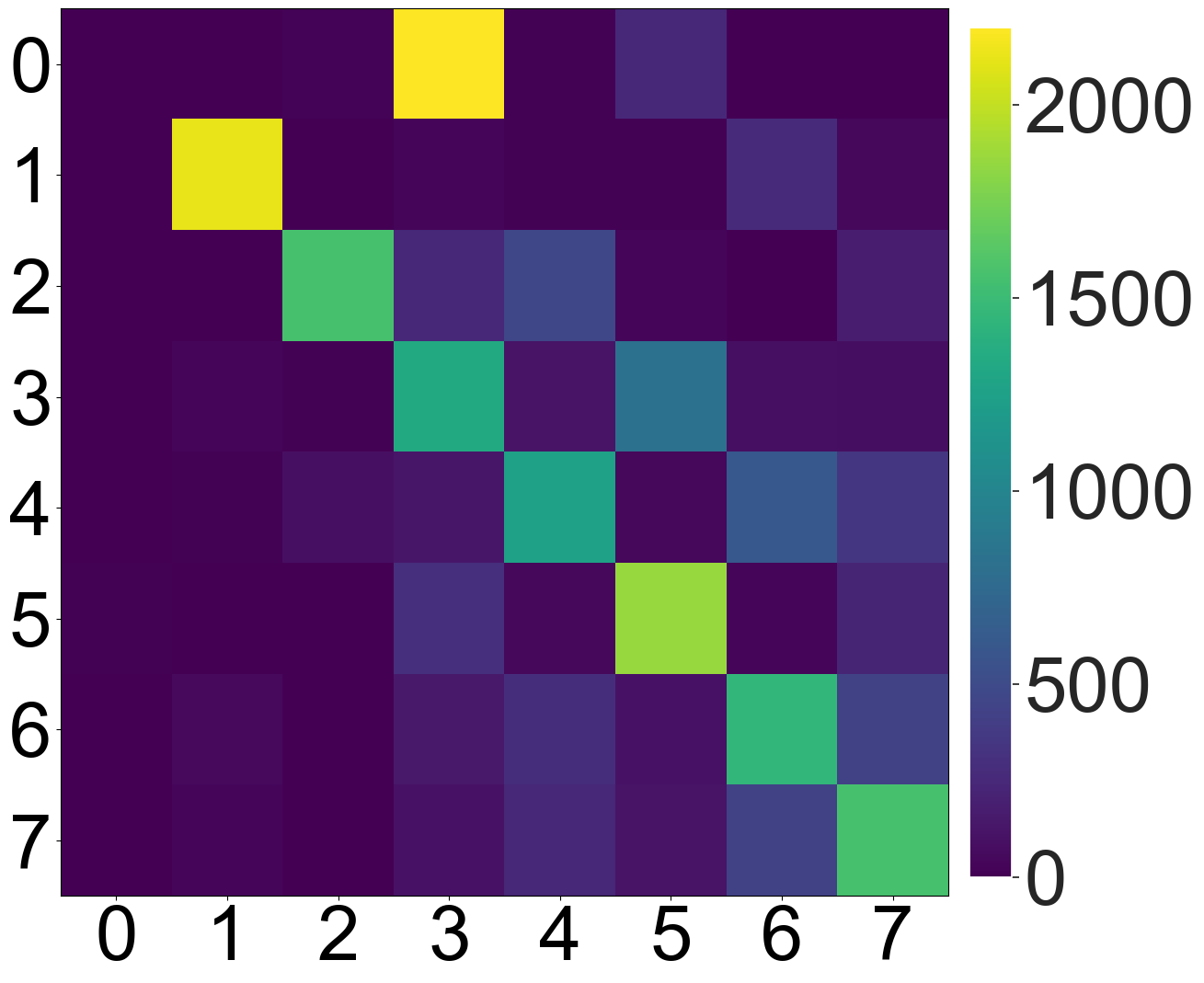}
  \caption{Model 3: $[200-200]$}
  \label{fig:3}
\end{subfigure}

\medskip
\begin{subfigure}{0.25\textwidth}
  \includegraphics[width=\linewidth]{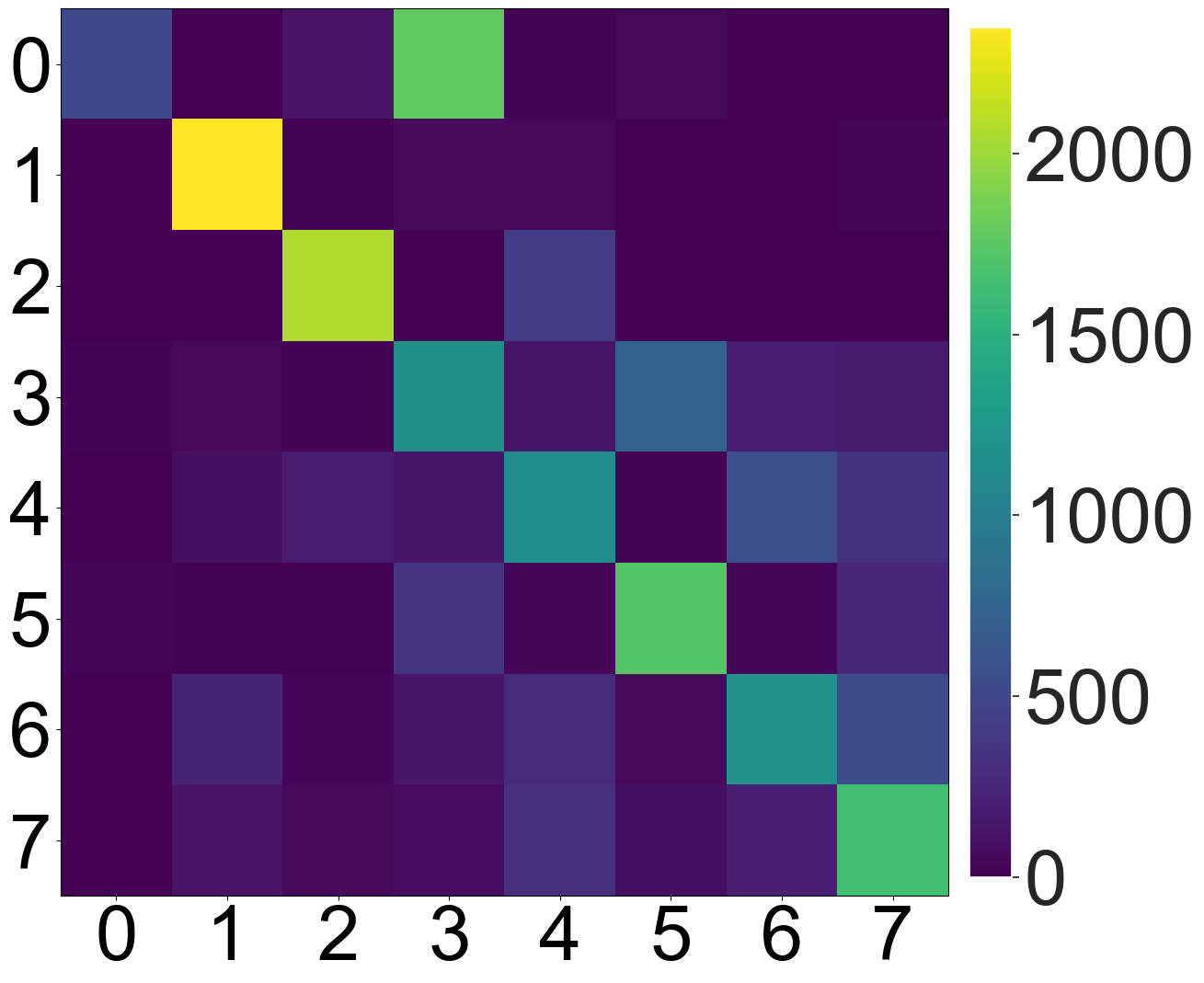}
  \caption{Model 4: $[100-200-300]$}
  \label{fig:4}
\end{subfigure}\hfil 
\begin{subfigure}{0.25\textwidth}
  \includegraphics[width=\linewidth]{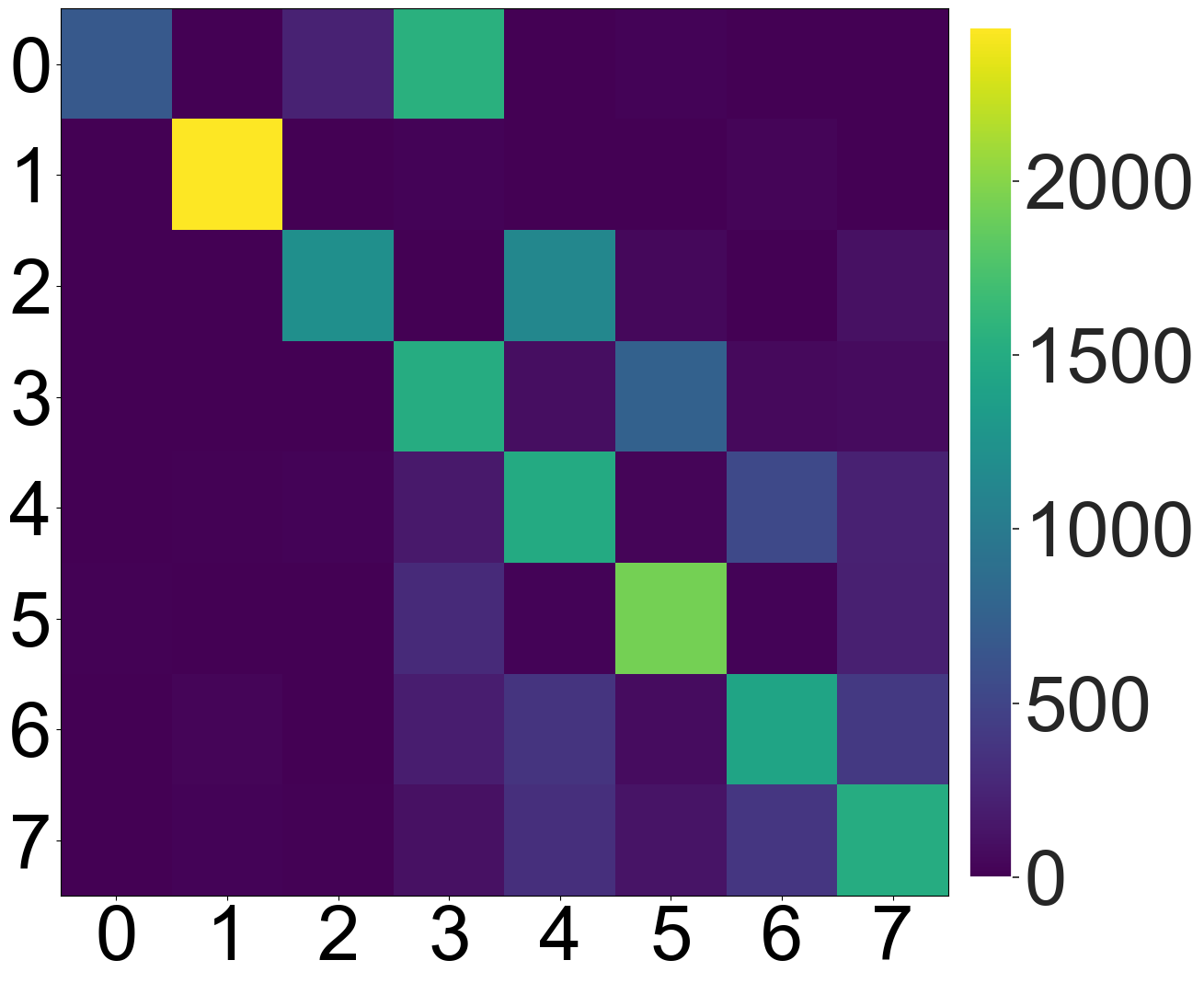}
  \caption{Model 5: $[300-200-100]$}
  \label{fig:5}
\end{subfigure}\hfil 
\begin{subfigure}{0.25\textwidth}
  \includegraphics[width=\linewidth]{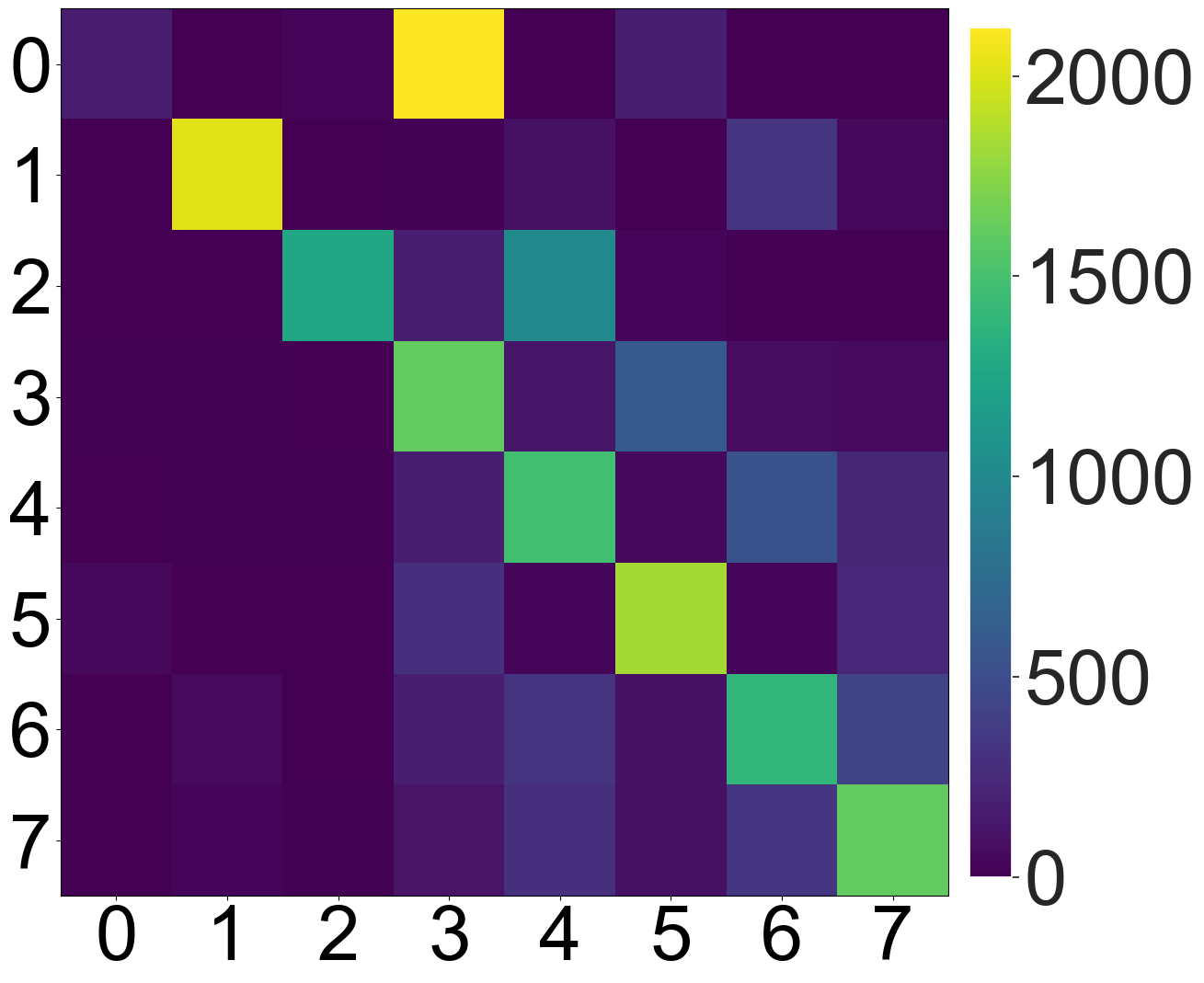}
  \caption{Model 6: $[300-300-300]$}
  \label{fig:6}
\end{subfigure}
\caption{Heat map of the confusion matricies of our trained architectures. The vertical axis corresponds to true labels and the horizontal axis to predicted labels. A perfect model would exhibit a diagonal confusion matrix.}
\label{fig:confusion_heatmap}
\end{figure*}

The confusion matrix is a valuable analysis tool for scrutinizing the performance of a model in detail. From the confusion matrix, we extracted the recall and precision of each class, as well as the overall accuracies and macro F1 scores of the models. All of these metrics range from $[0,1]$, with a higher score indicating better performance. In multiclass classification, recall and precision treat the correct class as ``positive'' while the other classes are ``negatives.'' Precision indicates how well a model correctly identifies a particular class, while recall reflects the model’s ability to identify all instances of a specific class. Accuracy quantifies the number of correctly predicted outcomes out of all classes. Additionally, the macro F1 score, the harmonic mean of the macro-averaged recalls and precisions, provides a balanced measure of a model's performance. For a review of these metrics, one could refer to \cite{kelleher2020fundamentals, grandini2020metrics}. We visualize the confusion matrix of the six models by plotting their heat maps, as shown in Figure~\ref{fig:confusion_heatmap}. The confusion matrices themselves can be accessed in our GitHub repository \cite{MyPhDGithubSantos}.

The vertical axis corresponds to true classes, and the horizontal axis represents predicted classes. In an ideal scenario, where the model perfectly predicts each class, the confusion matrix would form a diagonal pattern. However, we observe deviation towards the lower right end of the heat map. Since our dataset is organized in ascending complexity, our neural network struggled towards the end of the data sequence. The accuracies and F1 scores of the models are listed in Table~\ref{tab:model_summary}.
\begin{table}[tbhp!]
\caption{\label{tab:model_summary}Average performance metrics versus validation dataset across models.}
\begin{ruledtabular}
\begin{tabular}{ccc}
\textrm{Model} & \textrm{Accu.} & \textrm{Macro F1} \\
\colrule
$1$ & $0.59$ & $0.29$ \\
$2$ & $0.62$ & $0.32$ \\
$3$ & $0.56$ & $0.28$ \\
$4$ & $0.59$ & $0.30$ \\
$5$ & $0.61$ & $0.32$ \\
$6$ & $0.57$ & $0.30$ \\
\end{tabular}
\end{ruledtabular}
\end{table}

\begin{table*}[tbhp!]
    \centering
    \caption{\label{tab:modelscores} Labelwise performance of our models using a validation dataset.}
    \begin{tabular}{ccccc}
        \begin{tabular}{>{\centering\arraybackslash}m{1cm}>{\centering\arraybackslash}m{2cm}>{\centering\arraybackslash}m{1.5cm}}
            \toprule
            M1 & Precision & Recall  \\
            \colrule
            0 & 0.23 & 0.00   \\
            1 & 0.83 & 0.92   \\
            2 & 0.79 & 0.88   \\
            3 & 0.32 & 0.49   \\
            4 & 0.53 & 0.52   \\
            5 & 0.60 & 0.72   \\
            6 & 0.56 & 0.56   \\
            7 & 0.56 & 0.6   \\
            \toprule
        \end{tabular}
        \hspace*{4em}
        \begin{tabular}{>{\centering\arraybackslash}m{1cm}>{\centering\arraybackslash}m{2cm}>{\centering\arraybackslash}m{1.5cm}}
            \toprule
            M2 & Precision & Recall  \\
            \colrule
            0 & 0.89 & 0.41   \\
            1 & 0.84 & 0.94   \\
            2 & 0.89 & 0.70   \\
            3 & 0.39 & 0.50   \\
            4 & 0.48 & 0.52   \\
            5 & 0.63 & 0.71   \\
            6 & 0.54 & 0.53   \\
            7 & 0.56 & 0.63   \\
            \toprule
        \end{tabular}
        \hspace*{4em}
        \begin{tabular}{>{\centering\arraybackslash}m{1cm}>{\centering\arraybackslash}m{2cm}>{\centering\arraybackslash}m{1.5cm}}
            \toprule
            M3 & Precision & Recall  \\
            \colrule
            0 & 0.13 & 0.00   \\
            1 & 0.93 & 0.85   \\
            2 & 0.92 & 0.62   \\
            3 & 0.30 & 0.54   \\
            4 & 0.52 & 0.50   \\
            5 & 0.57 & 0.74   \\
            6 & 0.50 & 0.58   \\
            7 & 0.54 & 0.62   \\
            \toprule
        \end{tabular} \\
        
        \\
        \\
        
        \begin{tabular}{>{\centering\arraybackslash}m{1cm}>{\centering\arraybackslash}m{2cm}>{\centering\arraybackslash}m{1.5cm}}
            \toprule
            M4 & Precision & Recall  \\
            \colrule
            0 & 0.90 & 0.21   \\
            1 & 0.82 & 0.94   \\
            2 & 0.82 & 0.83   \\
            3 & 0.32 & 0.47   \\
            4 & 0.47 & 0.46   \\
            5 & 0.64 & 0.69   \\
            6 & 0.54 & 0.48   \\
            7 & 0.54 & 0.65   \\
            \toprule
        \end{tabular}
        \hspace*{4em}
        \begin{tabular}{>{\centering\arraybackslash}m{1cm}>{\centering\arraybackslash}m{2cm}>{\centering\arraybackslash}m{1.5cm}}
            \toprule
            M5 & Precision & Recall  \\
            \colrule
            0 & 0.98 & 0.27   \\
            1 & 0.97 & 0.98   \\
            2 & 0.81 & 0.49   \\
            3 & 0.39 & 0.60   \\
            4 & 0.44 & 0.60   \\
            5 & 0.64 & 0.77   \\
            6 & 0.58 & 0.57   \\
            7 & 0.60 & 0.60   \\
            \toprule
        \end{tabular}
        \hspace*{4em}
        \begin{tabular}{>{\centering\arraybackslash}m{1cm}>{\centering\arraybackslash}m{2cm}>{\centering\arraybackslash}m{1.5cm}}
            \toprule
            M6 & Precision & Recall  \\
            \colrule
            0 & 0.73 & 0.07   \\
            1 & 0.94 & 0.81   \\
            2 & 0.94 & 0.50   \\
            3 & 0.34 & 0.64   \\
            4 & 0.44 & 0.59   \\
            5 & 0.64 & 0.74   \\
            6 & 0.52 & 0.56   \\
            7 & 0.61 & 0.65   \\
            \toprule
        \end{tabular}
    \end{tabular}
\end{table*}

In machine learning, what qualifies as ``good accuracy'' depends on the problem or target goal for which the model is created. In our work, we are classifying eight classes, with a baseline threshold set to $1/8 = 0.125$. Since the accuracy scores listed in Table~\ref{tab:model_summary} have outperformed our baseline threshold, we may conclude that our models have performed decently. Alternatively, taking a more conservative approach, we could say that only models $2$ and $5$ achieved good accuracy scores of $0.62$ and $0.61$, respectively. It is important to note that accuracy only tells us how well a model predicts the correct classes in a given dataset; it does not indicate which classes our models may be less effective at identifying.

On the other hand, the macro F1 scores of our models are relatively low when compared to their accuracies. Given that this is a harmonic mean of the macro-averaged precision and recall and the discrepancy between it and the accuracy scores, it suggests where the models may be underperforming; either in recall or precision. We could examine the heat map in Figure~\ref{fig:confusion_heatmap} for clues and refer to the precision and recall scores in Table~\ref{tab:modelscores} for a more definitive answer.

Figure~\ref{fig:confusion_heatmap} shows that class $0$ is misidentified as class $3$, i.e., a single pole on the $[bt]$ sheet is seen as a single pole on the $[bt]$ sheet plus a single pole on the $[bb]$ sheet. We can understand this by comparing the line shape amplitude of a single pole on the $[bt]$ sheet with that on a $[bb]$ sheet. The pole on the $[bt]$ sheet dominates the pole on the $[bb]$ sheet below and at the $\Sigma_c\Bar{D}$ threshold. This misidentification of class $0$ as class $3$ is reflected through the low recall scores of class $0$ across all models. On the other hand, the precision scores of class $0$ are very good except for models $1$ and $3$. In other words, four models ($2$, $4$, $5$, and $6$) can discriminate other classes well from class $0$ -- the occurrence of false positives is low for these four models. Even though all our models cannot competitively identify class $0$ in the given $2,500$ instances, we are assured that four of them can distinguish other classes from class $0$. Therefore, in the event the inference results give us a prediction other than class $0$, then we can safely trust the model.

Similarly, low scores for recall and precision are also evident for class $3$. A good amount of instances of class $3$ are misidentified as class $5$, i.e., a configuration with one pole each on the $[bt]$ and $[bb]$ sheets is incorrectly identified as a configuration with a single pole on each of the unphysical sheets. This can be understood intuitively because a pole on the $[tb]$ sheet is far away from the physical region, and its contribution to the lineshape amplitude is minimal compared to the two poles on the $[bt]$ and $[bb]$ sheets. Due to the low recall and precision scores for class $3$, we cannot rely on the models to accurately predict this class.

Turning our attention to class $5$, the models sometimes tag class $3$ as class $5$. All recall scores for class $5$ are $0.70$ or higher, except for model $4$, which scored $0.69$. This implies that, at most, around $30\%$ of the validation dataset for class $5$ were misclassified as other classes. Conversely, its precision score hovers around $0.6$ on average for all six models. This means that around $60\%$ of the predicted class $5$ are correct, while the other $40\%$ were misclassifications from other classes.

Therefore, we can see why all of our models achieved low macro F1 scores. The precision and recall of class $0$ greatly affected it. Nevertheless, the low macro F1 scores in Table~\ref{tab:model_summary} should not hinder us from using our model to analyze the LHCb run 2 data. We can safely use it as long as we understand the context for which the models are created and, more importantly, what each precision and recall score for the classes means in the results of our inference stage.

\section{Application to the $J/\psi p$ scattering}\label{sec:IV}
Having trained the six neural networks, we transition to the inference stage. Ref.~\cite{SombilloPhysRevD.104.036001} utilized the uncertainty inherent in the experimental data to generate line shape amplitudes. We will delve into the mechanics of this process, outlining our approach to inference, and explore its implications in the subsequent subsections.
\subsection{Inference stage}
In constructing our training dataset, we introduced uncertainty to the particle detector by binning the energy axis. For our inference dataset, we reconstructed the experimental data by incorporating both the uncertainty along the energy axis and the weighted amplitude axis. The energy axis was constructed by randomly selecting $x_n$ points within the error bars of the experimental data's energy axis. Correspondingly, the value of $y_n(x_n)$ was chosen within the range $y_n \pm \text{unc.}_\pm$. For both the random selection of energy points and the corresponding amplitude values, we used a uniform probability distribution. To this end, we generated a total of $3,000$ samples of line shape amplitudes from the experimental data \cite{virtualLHCbcollab,LHCB2019hepdata.89271} to feed to the trained neural networks. We summarize our findings in Table~\ref{tab:inference}.
\begin{table}[tbhp!]
\centering
\caption{\label{tab:inference}Model list and their corresponding inference result.}
\begin{ruledtabular}
\begin{tabular}{ccdr}
\textrm{Model} & \textrm{Inference result} \\
\colrule
$1$ & class $5$: 1 pole each in $[bt]$, $[bb]$ and $[tb]$ \\
$2$ & class $5$: 1 pole each in $[bt]$, $[bb]$ and $[tb]$ \\
$3$ & class $5$: 1 pole each in $[bt]$, $[bb]$ and $[tb]$ \\
$4$ & class $3$: 1 pole each in $[bt]$ and $[bb]$ \\
$5$ & class $5$: 1 pole each in $[bt]$, $[bb]$ and $[tb]$ \\
$6$ & class $5$: 1 pole each in $[bt]$, $[bb]$ and $[tb]$ \\
\end{tabular}
\end{ruledtabular}
\end{table}

For each model, all $3,000$ samples were inferred with $100\%$ accuracy. Model $4$ inferred that the experimental data has a class $3$ pole structure. However, as discussed in the previous section, with class $3$ achieving low precision and recall scores across all models, this result is moot. Meanwhile, the rest of the models inferred that the experimental data has a class $5$ pole structure, i.e., it has a single pole on each of the unphysical sheets. Referring to Table~\ref{tab:modelscores}, the average precision and recall of class $5$ across all the five remaining models are $0.61$ and $0.74$, respectively. This means that whenever the models predict a positive class $5$, it is correct $61\%$ of the time and identifies positive instances of class $5$ $74\%$ of the time. If we were to use accuracy as our metric (see Table~\ref{tab:model_summary}), then the accuracy of this three-pole structure hovers around $60\%$. Lastly, we remark that when we switch from a uniform to a Gaussian probability distribution for constructing the inference dataset, the findings presented in Table~\ref{tab:inference} remain unchanged, even when accounting for deviations up to $5\sigma$.

\subsection{Discussion of results}
Majority of the DNN models that we considered favored the three pole structure of the $P_\psi^N(4312)^+$. Namely, one pole in each of the unphysical Riemann sheet within a coupled-two channel simplification. It is important to note that we utilized the off-diagonal element of the two-channel S-matrix in the generation of the training dataset in order to remove the possible ambiguity that will arise between the structure produced by an isolated pole in $[bt]$ sheet with that of the three-pole structure \cite{SantosSombilloPhysRevC.108.045204}. At this point, one can now attach a dynamical model to the obtained pole structure. 

An isolated pole in the fourth Riemann sheet can be produced by the combined effect of channel coupling and the weak $\Sigma_c\bar{D}$ attraction of the higher mass channel resulting to a virtual state in the zero coupling limit. Further increasing the channel coupling parameter may drag this pole from the fourth Riemann sheet to the second Riemann sheet by crossing the real energy axis above the second threshold \cite{Frazer1964,PearceGibson,Pc2019}. However, for moderate channel coupling, the pole stays at the fourth Riemann sheet. Now, if there is a compact state that decays into the lower channel $J/\psi p$, we can expect a pole and shadow pole pair in the second and third Riemann sheet. This pole structure is consistent with the pole counting argument that two poles in different Riemann sheet are non-molecular in nature \cite{Morgan1992,Baru2004}. This three pole structure can then produce a line shape that looks very similar to an isolated second Riemann sheet pole. This implies that the compact nature of $P_\psi^N(4312)^+$ is being contaminated by the coupling of virtual state produced in the $\Sigma_c\bar{D}$ channel to the lower $J/\psi p$ channel. 

The brute force line shape analysis via machine learning shows that the three pole structure is the most likely interpretation of the $P_\psi^N(4312)^+$. This interpretation is consistent with the hybrid model proposed in \cite{Yamaguchi:2017zmn, Yamaguchu2020} and supports the possibility of compact bound state analyses in \cite{ZHANG2023981,Strakovsky:2023kqu, Wang:2022ltr}.




\section{Conclusion and outlook}\label{sec:V}

Neural networks excel in recognizing patterns and generalizing information beyond the training dataset. To construct our training dataset, we used the independent $S$-matrix poles formulation to generate line shape amplitudes for the invariant mass spectrum of $J/\psi p$. By utilizing uniformization, we were able to rigorously account for proper near-threshold behaviors \cite{Yamada2020} and freely control the poles of the $S$-matrix without violating its constraints. Furthermore, this method ensures that there is no bias towards any specific trajectory when attaining a particular pole configuration \cite{PearceGibson, Frazer1964, Hanhart2014, Hyodo:2014bda}. With only unitarity (below the threshold), hermiticity, analyticity, and the dominance of $s$-waves due to the strong cusp at the $\Sigma_c^+\Bar{D}$ threshold as constraints, we argue that our generated dataset is minimally biased. Additionally, we included the off-diagonal $S_{12}$ term in the construction of our total line shape amplitude to distinguish ambiguous line shapes arising from the parametrization we used \cite{SantosSombilloPhysRevC.108.045204}.

Having constructed our training dataset with the stipulations above, we trained six neural networks to identify the pole structure of $P_\psi^N(4312)^+$. We implemented curriculum learning to help the networks converge faster to a minimum. The neural networks achieved decent accuracy, hovering around $0.60$, but attained low macro F1 scores. The low macro F1 scores were attributed to the poor performance of the neural network on class $0$ (one pole on $[bt]$) and class $3$ (one pole each on $[bt]$ and $[bb]$) in our model space.

Upon inference, five out of the six models indicated that the $P_\psi^N(4312)^+$ signal has a class $5$ pole structure, i.e., one pole each on the $[bt]$, $[bb]$, and $[tb]$ sheets. The result of the remaining one model is subject to dispute, as it inferred a class that attained low recall and precision scores. Overall, erring on the conservative side, the five models that inferred class $5$ have an average precision and recall scores of $0.61$ and $0.74$, respectively. 

However, we emphasize that our results are only indicative and should not be used as proof. Regardless, our result represents the first report of a plausible three-pole structure of $P_\psi^N(4312)^+$ and surely warrants future investigation. Moving forward, we strongly suggest that the contribution of the inelastic channel $S_{12}$ not be ignored and that a dynamic model study be conducted, building upon the three-pole structure we found.

On the side of machine learning, we recommend performing a grid search. In this work, we only considered several parameters, e.g., the number of layers and number of nodes within. This restricts the parameter space and may lock us from achieving a more robust model.

\section*{Acknowledgment}
We thank D.A.A. Co, J.R. Mabajen, and C. Villano for the discussions. We would also like to extend our gratitude to J. J. Operaña, K. T. Cervantes, C. Villano, and A.C.C. Alipio for helping us generate our dataset.






\newpage
\bibliography{mybib}

\end{document}